\documentclass[aip,jap,reprint]{revtex4-1}
\usepackage{graphicx}
\usepackage{amsmath}

\newcommand{\phidot}{\dot\phi}
\newcommand{\tr}{\mathop\mathrm{Tr}}
\newcommand{\im}{\mathop\mathrm{Im}}

\begin{document}

\title{Origin of the spin reorientation transitions in (Fe$_{1-x}$Co$_{x}$)$_{2}$B alloys}
\thanks{This article has been accepted by Applied Physics Letters. After it is published, it will be found at \url{http://scitation.aip.org/content/aip/journal/apl}.}
\author{Kirill D. Belashchenko}
\affiliation{Department of Physics and Astronomy and Nebraska Center for Materials and Nanoscience,
University of Nebraska-Lincoln, Lincoln, Nebraska 68588, USA}
\author{Liqin Ke}
\affiliation{Ames Laboratory, U.S. Department of Energy, Ames, Iowa 50011, USA}
\author{Markus D\"ane}
\affiliation{Lawrence Livermore National Laboratory, Livermore, California 94550, USA}
\author{Lorin X. Benedict}
\affiliation{Lawrence Livermore National Laboratory, Livermore, California 94550, USA}
\author{Tej Nath Lamichhane}
\affiliation{Ames Laboratory, U.S. Department of Energy, Ames, Iowa 50011, USA}
\affiliation{Department of Physics and Astronomy, Iowa State University, Ames, Iowa 50011, USA}
\author{Valentin Taufour}
\affiliation{Ames Laboratory, U.S. Department of Energy, Ames, Iowa 50011, USA}
\affiliation{Department of Physics and Astronomy, Iowa State University, Ames, Iowa 50011, USA}
\author{Anton Jesche}
\affiliation{Ames Laboratory, U.S. Department of Energy, Ames, Iowa 50011, USA}
\affiliation{Department of Physics and Astronomy, Iowa State University, Ames, Iowa 50011, USA}
\author{Sergey L. Bud'ko}
\affiliation{Ames Laboratory, U.S. Department of Energy, Ames, Iowa 50011, USA}
\affiliation{Department of Physics and Astronomy, Iowa State University, Ames, Iowa 50011, USA}
\author{Paul C. Canfield}
\affiliation{Ames Laboratory, U.S. Department of Energy, Ames, Iowa 50011, USA}
\affiliation{Department of Physics and Astronomy, Iowa State University, Ames, Iowa 50011, USA}
\author{Vladimir P. Antropov}
\affiliation{Ames Laboratory, U.S. Department of Energy, Ames, Iowa 50011, USA}

\date{\today}

\begin{abstract}
Low-temperature measurements of the magnetocrystalline anisotropy energy $K$ in (Fe$_{1-x}$Co$_{x}$)$_{2}$B alloys are reported, and the origin of this anisotropy is elucidated using a first-principles electronic structure analysis. The calculated concentration dependence $K(x)$ with a maximum near $x=0.3$ and a minimum near $x=0.8$ is in excellent agreement with experiment. This dependence is traced down to spin-orbital selection rules and the filling of electronic bands with increasing electronic concentration. At the optimal Co concentration, $K$ depends strongly on the tetragonality and doubles under a modest 3\% increase of the $c/a$ ratio, suggesting that the magnetocrystalline anisotropy can be further enhanced using epitaxial or chemical strain.
\end{abstract}

\maketitle

Magnetocrystalline anisotropy (MCA) of a magnetic material is one of its key properties for practical applications, large easy-axis anisotropy being favorable for permanent magnets. \cite{STOHR} Intelligent search for new materials requires understanding of the underlying mechanisms of MCA. This can be particularly fruitful for substitutional alloys whose properties can be tuned by varying the concentrations of their components.
The analysis is often relatively simple in insulators, where MCA is dominated by single-ion terms which can be deduced from crystal-field splittings and spin-orbital (SO) selection rules. In contrast, in typical metallic alloys the band width sets the largest energy scale, and MCA depends on the details of the electronic structure.

The (Fe$_{1-x}$Co$_{x}$)$_{2}$B solid solution \cite{IGA,Cadeville,TAK,COENE,KUZ} (space group $I4/mcm$\cite{structure}) is a remarkable case in point. Fe$_{2}$B has a fairly strong easy-plane MCA, and Co$_{2}$B, at low temperatures, a small easy-axis MCA. However, the alloy has a substantial easy-axis MCA around $x=0.3$, \cite{IGA} making it a potentially useful rare-earth-free \cite{Lewis} permanent magnet. At $x\approx0.5$ the MCA again turns easy-plane, peaks at $x=0.8$, and then turns easy-axis close to $x=1$. These three spin reorientation transitions must be related to the continuous evolution of the electronic structure with concentration. The goal of this Letter is to elucidate the origin of this rare phenomenon.

\begin{figure}[htb]
\includegraphics[width=0.97\columnwidth]{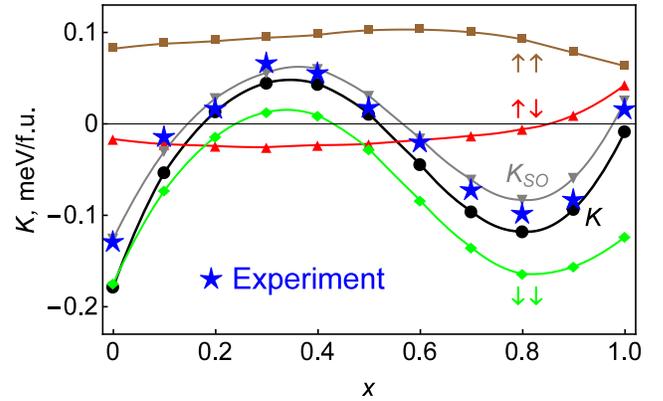}
\caption{Calculated (black circles) and experimental (blue stars) MCA energy $K$ in (Fe$_{1-x}$Co$_x$)$_2$B alloys. Gray curve: $K_{SO}$. The other lines show the spin decomposition $K_{\sigma\sigma'}$.}
\label{MAEfig}
\end{figure}

First, we report the results of experimental measurements at low temperatures. Single crystals of (Fe$_{1-x}$Co$_x$)$_2$B were grown from solution growth out of an excess of (Fe,Co) which was decanted in a centrifuge. \cite{CF} The single crystals were grown as tetragonal rods which were cut using a wire saw to give them the shape of a rectangular prism. The demagnetization factor was calculated using Ref.\ \onlinecite{Aharoni}. Field-dependent magnetization measurements were performed in a Quantum Design MPMS at 2 K in fields up to 5.5 T. The MCA energy $K$ was determined as the area between the two magnetization curves, with the field parallel and perpendicular to the $c$ axis, taken at the same temperature. \cite{KUZ} The results shown in Fig.\ \ref{MAEfig} (blue stars) measured at 2 K confirm the nonmonotonic concentration dependence, in good agreement with the measurements at 77 K from Ref.\ \onlinecite{IGA}.

Density-functional calculations using several different methods show that the choice of the exchange-correlation potential and other computational details strongly affect the calculated MCA in Fe$_2$B and Co$_2$B. We have ascertained that this sensitivity is largely due to the variation of the exchange splitting, which controls the position of the Fermi level relative to the minority-spin bands. The systematic variation of MCA with $x$ is also controlled by the Fermi level shift. This continuous variation is, therefore, reliably predicted as long as the end points are correctly fixed using experimental input.

The results reported below were obtained using the Green's function-based formulation of the tight-binding linear muffin-tin orbital (GF-LMTO) method. \cite{Turek} Substitutional disorder was treated using our implementation of the coherent potential approximation (CPA),\cite{Fe8N} with the SO coupling included perturbatively as described in the supplementary material. \cite{supplement}

The lattice constants and the internal coordinates are linearly interpolated between the experimental data for the end compounds extrapolated to zero temperature: \cite{IGA} $a=5.109$ and 4.997 \AA, $c=4.249$ and 4.213 \AA, and $u=0.166$ and 0.168 for Fe$_2$B and Co$_2$B, respectively. The exchange and correlation are treated within the generalized gradient approximation (GGA). \cite{PBE}

\begin{figure}[htb]
\includegraphics[width=0.97\columnwidth]{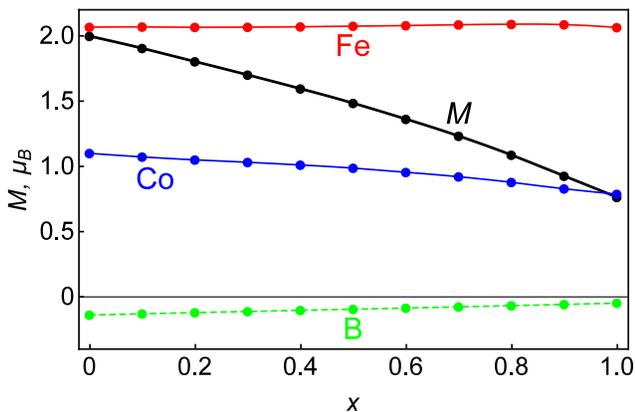}
\caption{Spin moments on different atoms and the total spin magnetization $M$ per transition-metal atom.}
\label{moments}
\end{figure}

The correct exchange splitting at the end points can be enforced by using the experimental data \cite{Cadeville,TAK,COENE,IGA,KUZ} for the magnetization $M$: 1.9 $\mu_B$/Fe in Fe$_2$B and 0.76 $\mu_B$/Co in Co$_2$B. In Fe$_2$B it is only slightly overestimated, but in Co$_2$B it is much too large at about 1.1 $\mu_B$/Co in all density-functional calculations. The relatively small spin moment of Co indicates a pronounced itinerant character of magnetism in Co$_2$B, which tends to be sensitive to quantum spin fluctuations. Therefore, we introduced a scaling factor for the local part of the effective magnetic field in the GGA functional for the Co atoms and adjusted it to match the experimental magnetization. The resulting scaling factor of 0.80 was then used for Co atoms at all concentrations. The spin moments on different atoms and the total spin magnetization obtained in this way are shown in Fig.\ \ref{moments}. While the Fe spin moment is almost constant, the Co spin moment declines with $x$. Moreover, this decline accelerates at $x\gtrsim0.6$, which is in excellent agreement with experimental data. \cite{Cadeville}

The MCA energy $K$ was obtained by calculating the single-particle energy difference for in-plane and out-of-plane orientations of the magnetization while keeping the LMTO charges fixed at their self-consistent values found without SO coupling. A uniform mesh of 30$^3$ points in the full Brillouin zone provided sufficient accuracy for the $\mathbf{k}$ integration. We have verified that the values of $K$ for pure Fe$_2$B and Co$_2$B agree very well with Hamiltonian LMTO results. The concentration dependence of $K$ is shown in Fig.\ \ref{MAEfig}. The agreement with low-temperature experimental data is remarkably good, suggesting that the electronic mechanisms of MCA are correctly captured in the calculations. If the spin moment of Co is not corrected by scaling the exchange-correlation field, the downward trend in $K$ at the Co-rich end continues to large negative values in disagreement with experiment.

In $3d$ systems the SO band shifts are usually well described by second-order perturbation theory, except perhaps in small regions of the Brillouin zone.
Consequently, when MCA appears in second order in SO coupling (as in the tetragonal system under consideration), the anisotropy of the expectation value of the SO operator $\Delta E_{SO}=\langle V_{SO}\rangle_x - \langle V_{SO}\rangle_z$ is approximately equal to $2K$. \cite{supplement} (Here $x$ or $z$ shows the orientation of the magnetization axis.) We therefore denote $K_{SO}=\Delta E_{SO}/2$ and use the expression $2\langle\mathbf{SL}\rangle=\langle L_{z'}\rangle_{\uparrow\uparrow}-\langle L_{z'}\rangle_{\downarrow\downarrow}+\langle L_+\rangle_{\downarrow\uparrow}+\langle L_-\rangle_{\uparrow\downarrow}$ to separate the contributions to $K$ by pairs of spin channels. Here we use $L_{z'}$ to denote the component of $\mathbf{L}$ parallel to the magnetization axis, to avoid confusion with the crystallographic $z$ axis; $L_\pm$ are the usual linear combinations of the other two (primed) components of $\mathbf{L}$. The contributions $K_{\sigma\sigma'}$ to $K_{SO}$ are accumulated as energy integrals taking into account the energy dependence of the SO coupling parameters. The results for $K_{SO}$ and $K_{\sigma\sigma'}$ are shown in Fig.\ \ref{MAEfig}. First, we see that $K_{SO}$ is close to $K$, confirming the validity of this analysis. Second, the nonmonotonic concentration dependence of $K$ is almost entirely due to $K_{\downarrow\downarrow}$ for $x\lesssim0.7$. While $K_{\uparrow\uparrow}$ is sizeable, it depends weakly on $x$ in this region. Additional discussion about the spin decomposition of MCA is provided in the supplementary material.\cite{supplement}

\begin{figure*}[htb]
\includegraphics[width=0.97\textwidth]{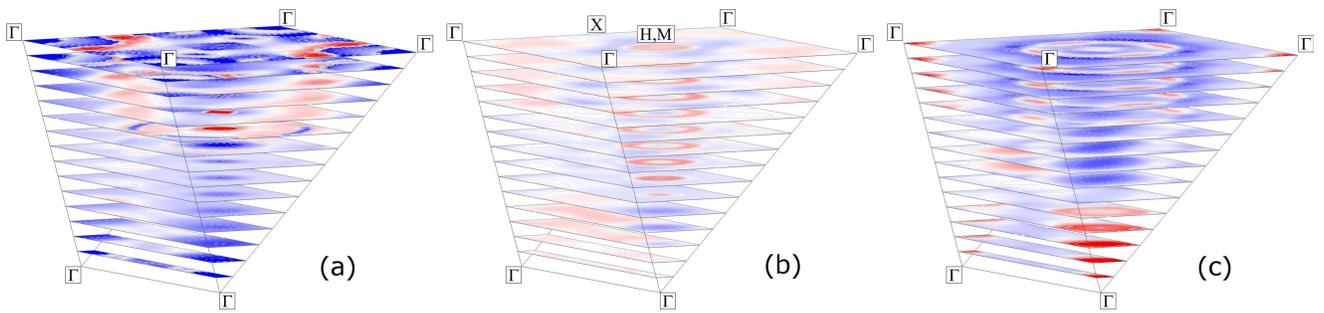}
\caption{Brillouin zone map of the $\mathbf{k}$-resolved minority-spin contribution to $K$ at: (a) $x=0$, (b) $x=0.3$, (c) $x=0.8$. Half of the Brillouin zone is shown; the top face of the plot is $k_z=0$. Points H (same as M) and X are shown in panel (b). The color intensity indicates the magnitude of negative (blue) and positive (red) values.}
\label{fig:bz}
\end{figure*}

Let us now analyze the electronic structure and the details of SO coupling-induced mixing for the minority-spin electrons. To resolve the minority-spin contribution to $K$ by wave vector $\mathbf{k}$, we calculated the minority-spin spectral function in the presence of SO coupling and found its first energy moment at each $\mathbf{k}$. Fig.\ \ref{fig:bz} shows the difference of these integrals for magnetization along the $x$ and $z$ axes at three key concentrations: pure Fe$_2$B ($x=0$), the maximum of $K$ ($x=0.3$) and its minimum ($x=0.8$). We have checked that the Brillouin zone integral of the $\mathbf{k}$-resolved MCA energy (summed up over both spins) agrees almost exactly with the value of $K$ calculated in the usual way.

Fig.\ \ref{fig:bz} shows that the MCA energy accumulates over a fairly large part of the Brillouin zone. At $x=0$ negative contributions to $K$ dominate over most of the Brillouin zone. At $x=0.3$ both positive and negative contributions are small. At $x=0.8$ there are regions with large positive and large negative contributions. Overall, it appears that the most important contributions come from the vicinity of the $\Gamma$XM ($k_z=0$) plane and from the vicinity of the $\Gamma$H ($k_x=k_y=0$) line.

\begin{figure*}[htb]
\includegraphics[width=0.97\textwidth]{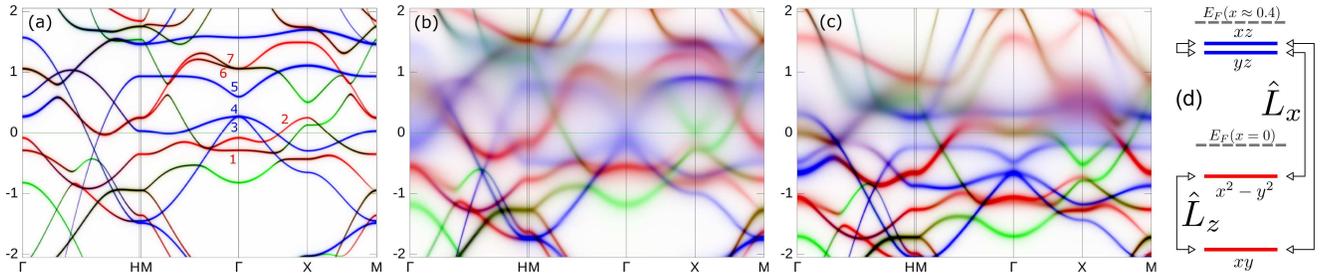}
\caption{(a-c) Minority-spin partial spectral functions for the transition-metal site in the absence of SO coupling at (a) $x=0$, (b) $x=0.3$, and (c) $x=0.8$. Energy is in eV. (d) Level diagram and SO selection rules at the $\Gamma$ point (bands 1-4). Color encodes the orbital character of the states. The intensities of the red, blue and green color channels are proportional to the sum of $xy$ and $x^2-y^2$ ($m=\pm2$), sum of $xz$ and $yz$ ($m=\pm1$), and $z^2$ ($m=0$) character, respectively.}
\label{fig:sf}
\end{figure*}

The partial minority-spin spectral functions for the transition-metal site are displayed in Fig.\ \ref{fig:sf} (panels (a)-(c)) along the important high-symmetry directions for the same three concentrations. The coloring in this figure resolves the contributions from different $3d$ orbitals. At $x=0$ the spectral function resolves the conventional electronic bands of pure Fe$_2$B (an imaginary part of 0.004 eV is added to the energy to acquire them). At $x=0.3$ and $x=0.8$ substitutional disorder broadens the bands by a few tenths of an electronvolt, but their identity is in most cases preserved. Thus, we will discuss the SO-induced band mixing in the alloy, bearing in mind that band broadening should reduce the values of MCA at intermediate $x$.

As we have learned above, the dominant concentration dependence of $K$ comes from the $\langle L_{z'}\rangle_{\downarrow\downarrow}$ term, where $z'$ is the magnetization axis. The electronic states on the whole $\Gamma$XM plane can be classified as even or odd under reflection $z \to -z$. Even (odd) states have $m=0,\pm2$ ($m=\pm1$) character and appear red and green (blue) in Fig.\ \ref{fig:sf}. States of different parity do not intermix on this plane in the absence of SO coupling, as is clearly seen in Fig.\ \ref{fig:sf}a. The selection rules for SO coupling of the minority-spin states follow from the definite parity of the components of $\hat{\mathbf{L}}$ under reflection. $\hat L_z$ (even; relevant for $\mathbf{M}\parallel z$) only mixes states of the same parity on the $\Gamma$XM plane, or more generally orbitals of the same $m$. Coupling between states of the $m=\pm2$ character (red) is stronger compared to states of the $m=\pm1$ character (blue). In contrast, $\hat L_x$ (odd; relevant for $\mathbf{M}\parallel x$) couples states of the opposite parity on the $\Gamma$XM plane, or more generally orbitals $m$ and $m\pm1$. All these couplings contribute to $K$ only when the Fermi level lies between the two states that are being coupled. Whenever $\hat L_z$ or $\hat L_x$ couples such states, there is a negative contribution to the energy of the system with the corresponding direction of $\mathbf{M}$.

With the help of Fig.\ \ref{fig:sf} we can now deduce which couplings contribute to $K$ at different concentrations. At $x=0$ the Fermi level lies between the filled even states (bands 1-2) and empty odd states (bands 3-4) near $\Gamma$. Coupling of these states by $\hat L_x$ contributes to negative $K$. Filling of the hole pocket at $\Gamma$ (bands 3-4) with increasing $x$ suppresses this contribution. At $x=0.3$ the odd bands 3-4 are filled (Fig.\ \ref{fig:sf}b), and the minority-spin contribution to $K$ is small (Fig.\ \ref{fig:bz}b). These two cases are sketched in Fig.\ \ref{fig:sf}d.

At still larger $x$ the odd band 5 gets gradually filled, activating the negative contribution to $K$ from the mixing of band 5 with empty even bands 6-7. This trend continues till about $x=0.8$, where an even pair of bands 6-7 (degenerate at $\Gamma$) begins to fill (Fig.\ \ref{fig:sf}c). Mixing of these bands by $\hat L_z$ leads to an intense positive contribution to $K$ near the $\Gamma$ point (clearly seen in Fig.\ \ref{fig:bz}c), and the trend reverses again. Thus, the nonmonotonic dependence of $K$ in the whole concentration range is explained. Note that if the exchange-correlation field on the Co atoms is not scaled down to bring the magnetization in agreement with experiment, the exchange splitting remains too large, and bands 6-7 remain empty up to $x=1$. As a result, without this correction the uninterrupted negative trend brings $K$ to large negative values in disagreement with experiment.

To assess the effect of atomic relaxations on $K$, we optimized all inequivalent structures with two formula units per unit cell using the VASP code \cite{vasp} and the GGA. The volumes were fixed at the same values as in the CPA studies at the same $x$, while the cell shape and internal coordinates were relaxed. Since all these supercells preserve the $\sigma_z$ reflection plane, the displacements of Fe and Co atoms are confined to the $xy$ plane. All displacements were less than 0.025 \AA. One of the two structures at $x=0.5$ breaks the $C_4$ symmetry. For this structure we took the average of the energies for $\mathbf{M}\parallel x$ and $\mathbf{M}\parallel y$ as the in-plane value in the calculation of $K$, which corresponds to the averaging over different orientations of the same local ordering.

The changes in the absolute value of $K$ due to the relaxation and its values (meV/f.u.) in the relaxed structures were found to be: $-26$\% and $-0.11$ in Fe$_2$B, $-6$\% and 0.25 in Fe$_{1.5}$Co$_{0.5}$B, $-13$\% and 0.15 in the FeCoB [100] superlattice, 6\% and 0.12 in the FeCoB [110] superlattice, $-11$\% and $-0.31$ in Fe$_{0.5}$Co$_{1.5}$B), and $-19$\% and $-0.04$ in Co$_2$B. While MCA tends to be larger for ordered structures, the concentration trend in supercell calculations agrees well with the CPA results for disordered alloys. Although this set of unit cells is limited, the results suggest that local relaxations do not qualitatively change the concentration dependence shown in Fig.\ \ref{MAEfig}.

\begin{figure}[htb]
\includegraphics[width=0.97\columnwidth]{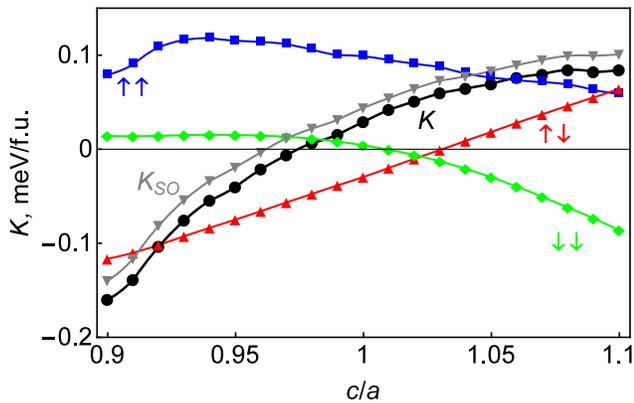}
\caption{MCA energy $K$ as a function of the $c/a$ ratio at $x=0.3$ (the value $c/a=1$ is assigned to the unstrained lattice). $K_{SO}$ and $K_{\sigma\sigma'}$ are also shown.}
\label{fig:kcoa}
\end{figure}

Larger positive values of $K$ are favorable for permanent magnet applications. Our electronic structure analysis shows that the maximum near $x\approx0.3$ corresponds to the optimal band filling. Further raising of $K$ requires favorable changes in the band structure, which could be induced by epitaxial or chemical strain. We therefore considered the dependence of $K$ at $x=0.3$ on the volume-conserving tetragonal distortion. The results plotted in Fig.\ \ref{fig:kcoa} show a very strong effect: $K$ is doubled under a modest 3\% increase in $c/a$ due to the sharply increasing spin-flip contributions. A more detailed analysis shows that the latter is largely due to the increase in the $c$ parameter. On the other hand, the minority-spin contribution increases with decreasing volume. Thus, increasing $c$ and decreasing $a$ both have a positive effect on MCA.
This enhancement could be achieved through epitaxial multilayer engineering. The search for a suitable alloying element to enhance the $c/a$ ratio is an interesting subject for further investigation.

In conclusion, we have explained how the spin reorientation transitions in (Fe$_{1-x}$Co$_{x}$)$_{2}$B alloys originate in the SO selection rules and the consecutive filling of minority-spin electronic bands of particular orbital character. Near the optimal 30\% concentration of Co, the MCA energy is predicted to increase quickly with the $c/a$ ratio, which could be implemented by epitaxial strain or a suitable chemical doping.

The work at UNL was supported by the National Science Foundation through Grant No.\ DMR-1308751 and performed utilizing the Holland Computing Center of the University of Nebraska. Work at Ames Lab and LLNL was supported in part by the Critical Materials Institute, an Energy Innovation Hub funded by the US DOE and by the Office of Basic Energy Science, Division of Materials Science and Engineering. Ames Laboratory is operated for the US DOE by Iowa State University under Contract No.\ DE-AC02-07CH11358. Lawrence Livermore National Laboratory is operated for the US DOE under Contract DE-AC52-07NA27344.

\pagebreak
\widetext
\begin{flushleft}
\textbf{\Large Supplemental Material:\\ Origin of the spin reorientation transitions in (Fe$_{1-x}$Co$_{x}$)$_{2}$B alloys}
\end{flushleft}

\setcounter{equation}{0}
\setcounter{figure}{0}
\setcounter{table}{0}
\setcounter{page}{1}
\makeatletter
\renewcommand{\thefigure}{S\arabic{figure}}
\makeatother

\section{Implementation of spin-orbit coupling in Green's function-based linear muffin-tin orbital (LMTO) method}

Magnetocrystalline anisotropy (MCA) of substitutional alloys is often studied using the fully-relativistic Korringa-Kohn-Rostocker or Green's function-based LMTO (GF-LMTO) method combined with the coherent potential approximation (CPA). \cite{Razee,Zabloudil,Turek2012,Weinberger} Here we employed a perturbative implementation of spin-orbit coupling in GF-LMTO. Dealing with conventional spinor wave functions, it simplifies the analysis of MCA without a significant loss of accuracy compared to the fully-relativistic approach.

\subsection{Relation between the Hamiltonian and Green's function-based LMTO formulation}

The so-called orthogonal Hamiltonian and Green's function-based LMTO formulations are equivalent to second order in $(E-E_\nu)$ where $E_{\nu}$ is the linearization energy. \cite{Gunnarsson,S-Turek} The Hamiltonian in the orthogonal basis is
\begin{equation}
H_{orth}=C+\sqrt{\Delta}S(1-\gamma S)^{-1}\sqrt{\Delta} , \label{H}
\end{equation}
where $C$, $\Delta$, and $\gamma$ are the diagonal matrices containing the LMTO potential parameters, and $S$ is the structure constant matrix. On the other hand, in GF-LMTO the Green's function
\begin{equation}
G(z)=\lambda(z)+\mu(z)[P(z)-S]^{-1}\mu(z),
\label{GF}
\end{equation}
is constructed using the diagonal matrices containing the potential functions
\begin{equation}
P(z)=\frac{z-C}{\Delta_l+\gamma_l(z-C)}, \quad
\mu(z)=\frac{\sqrt{\Delta}}{\Delta+\gamma(z-C)},\quad
\lambda(z)=\frac{\gamma}{\Delta+\gamma(z-C)}. \label{param}
\end{equation}
These definitions are related through energy derivatives (denoted by an overdot):
\begin{equation}
\mu^2=\dot P, \quad \lambda=-\frac12\ddot P/\dot P . \label{deriv}
\end{equation}
These relations guarantee that $G(z)$ does not have poles at the points where $P(z)$ has simple poles.
It is straightforward to show that $G(z)=(z-H_{orth})^{-1}$, which means that $G(z)$ in (\ref{GF}) is the exact resolvent of $H_{orth}$.

Third-order parametrization of the potential functions \cite{Gunnarsson} is often considerably more accurate compared to the second-order one. The accuracy of this parametrization is similar to that of the three-center LMTO Hamiltonian, although they are not exactly equivalent. In the nearly orthogonal LMTO basis, the Hamiltonian and overlap matrices are \cite{Andersen}
\begin{align}
H&=H_{orth} + (H_{orth}-E_\nu)E_\nu p (H_{orth}-E_\nu),\nonumber\\
O&=1+(H_{orth}-E_\nu)p(H_{orth}-E_\nu)
\end{align}
where $E_\nu$ and $p$ are diagonal matrices containing the linearization energies and the LMTO parameters $p=\langle\dot\phi^2\rangle$, and $H_{orth}$ is the same as in (\ref{H}). The generalized eigenvalue problem
is then written as
\begin{equation}
\det\left[(H_{orth}-E) - (H_{orth}-E_\nu)(E-E_\nu)p(H_{orth}-E_\nu)\right]=0
\end{equation}
The third-order term can be treated as a perturbation, and the variational correction formally amounts to the substitution $E\hat 1 \to E\hat 1 + (E-E_\nu)^{3} p$ in the second-order eigenvalue equation $\det(H_{orth}-E)=0$.

In the GF-LMTO formulation, third-order accuracy is similarly achieved by redefining the potential functions:\cite{Gunnarsson,Andersen}
\begin{align}
\tilde P = P(\tilde z), \quad
\tilde \mu =\sqrt{\tilde z^\prime}\mu(\tilde z), \quad
\tilde \lambda=-\frac12 \frac{\tilde z^{\prime\prime}}{\tilde z^\prime}+
\tilde z^\prime\lambda(\tilde z)\label{p3}
\end{align}
where $\tilde z = z+p(z-E_{\nu})^3$ and $\tilde z'=d\tilde z/dz$. These definitions are designed to preserve the relations (\ref{deriv}). If the third-order Green's function $\tilde G$ is defined similarly to (\ref{GF}) but with the third-order potential functions (\ref{p3}), it can be shown that
\begin{equation}
\tilde G = -\frac12 \frac{\tilde z^{\prime\prime}}{\tilde z^\prime} + \sqrt{\tilde z^\prime}(\tilde z - H_{orth})^{-1}\sqrt{\tilde z^\prime} . \label{G3}
\end{equation}

A solution of $\det (H_{orth}-\tilde z)=0$ corresponds to a pole of $\tilde G$, and the factors $\sqrt{\tilde z^\prime}$ guarantee that the residue at that pole is equal to 1. Thus, the eigenstates of the three-center LMTO Hamiltonian are correctly represented by $\tilde G$ to third order in $E-E_\nu$. The first term in (\ref{G3}) introduces two unphysical poles at $z=E_\nu \pm i/\sqrt{3p}$ for each orbital. It is a diagonal matrix which is real on the real axis and has no effect on the spectrum of the physical states.

\subsection{Spin-orbit coupling in GF-LMTO} \label{pert2}

In the Green's function formalism the perturbation can be introduced through the Dyson equation
\begin{equation}
G_V=G_0 + G_0\Sigma G_V \label{Dyson}
\end{equation}
where $G_0$ and $G_V$ are the unperturbed and perturbed Green's functions and $\Sigma$ the energy-dependent self-energy due to the perturbation $V$. For $V$ representing the spin-orbit coupling, $\Sigma$ is local.

In the second-order representation $G_0 = (z-H_{orth})^{-1}$, and therefore $G_V=(z-H_{orth}-\Sigma)^{-1}$. This $G_V$ can be constructed from the potential functions by denoting
\begin{equation}
C_V=C+\Sigma \label{CU}
\end{equation}
and defining the following nondiagonal matrices:
\begin{align}
P_V&=\sqrt{\Delta}[\Delta+(z-C_V)\gamma]^{-1}(z-C_V) \frac{1}{\sqrt{\Delta}}=\left[\gamma+\sqrt{\Delta}(z-C_V)^{-1}\sqrt{\Delta}\right]^{-1}\nonumber\\
\mu^L_V &=[\Delta+\gamma(z-C_V)]^{-1} \sqrt{\Delta} = (z-C_V)^{-1} \sqrt{\Delta} P_V\nonumber\\
\mu^R_V &=\sqrt{\Delta} [\Delta+(z-C_V)\gamma]^{-1} = P_V \sqrt{\Delta} (z-C_V)^{-1}\nonumber\\
\lambda_V &=[\Delta+\gamma(z-C_V)]^{-1}\gamma = \gamma[\Delta+(z-C_V)\gamma]^{-1}\label{PV}
\end{align}
The relations (\ref{deriv}) are then replaced by
\begin{equation}
\dot P_V = \mu_V^R \mu_V^L, \quad
\lambda_V =-\frac12 (\mu_V^R)^{-1}\ddot P (\mu_V^L)^{-1} . \label{pdot}
\end{equation}
A tedious but straightforward derivation then shows
\begin{equation}
G_V=\lambda_V+\mu_V^L(P_V-S)^{-1}\mu_V^R . \label{GU}
\end{equation}

If the unperturbed Green's function is calculated to third-order accuracy, we can still proceed from Dyson's equation, but the situation is complicated by the first term $\kappa=-\tilde z^{\prime\prime}/(2\tilde z^{\prime})$ in Eq.\ (\ref{G3}) for $\tilde G_0$, which modifies the perturbed Green's function $\tilde G_V$. However, for spin-orbit coupling this modification may be neglected, as we now argue. Let us denote $\tilde G_0=\kappa+\bar G_0$ and write down the diagrammatic expansion of $\tilde G_V$ in powers of $\Sigma$. It is just a usual expansion in which each Green's function line can be either $\bar G_0$ or $\kappa$. Now let us resum all diagram insertions with no external lines and no $\bar G_0$ lines inside. This gives a renormalized self-energy $\bar \Sigma =(1-\Sigma\kappa)^{-1}\Sigma$. Taking this into account and performing the remaining summations, we find
\begin{equation}
\tilde G_V = \bar G + \kappa + \kappa\bar\Sigma\kappa + \kappa\bar\Sigma\bar G + \bar G \bar\Sigma \kappa
\end{equation}
where
\begin{equation}
\bar G = (1-\bar G_0 \bar\Sigma)^{-1}\bar G_0.\label{barG}
\end{equation}
The quantity $\Sigma\kappa\sim p(E-E_\nu)\Sigma$ is very small for spin-orbit coupling in transition metals: $\Sigma\sim$ 50 meV, $p\sim 0.01$ eV$^{-2}$, $E-E_\nu\lesssim 5$ eV.  Since $\kappa$ is analytic on the real axis, $\bar\Sigma$ has no poles there too. Therefore, the poles of $\tilde G_V$ coincide with the poles of $\bar G$. The latter are just the poles of $\bar G_0$ which are shifted by the self-energy $\bar\Sigma$. But $\bar\Sigma$ differs from $\Sigma$ by a factor $(1-\Sigma\kappa)^{-1}$ which is equal to 1 up to small corrections of order $\Sigma\kappa$. Up to similar corrections, the residues of the poles of $\tilde G_V$ are equal to those for $\bar G$. Thus, neglecting all small terms or order $\Sigma\kappa$ compared to 1, we find $\tilde G_V\approx \bar G + \kappa$.

Substituting the second term in (\ref{G3}) as $\bar G_0$ in (\ref{barG}), we find the perturbed Green's function:
\begin{equation}
\tilde G_V = \sqrt{\tilde z^\prime}\left(\tilde z - H_{orth} - \sqrt{\tilde z^\prime} \Sigma \sqrt{\tilde z^\prime}\right)^{-1}\sqrt{\tilde z^\prime}  .
\end{equation}
where we have dropped the inconsequential term $\kappa$. This expression for $\tilde G_V$ is equivalent to
\begin{equation}
\tilde G_V=\tilde\lambda_V+\tilde\mu_V^L(\tilde P_V-S)^{-1}\tilde\mu_V^R \label{tGU}
\end{equation}
with redefined potential functions
\begin{equation}
\tilde P_V = P_V(\tilde z), \quad
\tilde \mu^L_V = \sqrt{\tilde z^\prime}\mu^L_V(\tilde z), \quad
\tilde \mu^R_V = \mu_V^R(\tilde z)\sqrt{\tilde z^\prime}, \quad
\tilde \lambda_V = \sqrt{\tilde z^\prime}\lambda_V(\tilde z)\sqrt{\tilde z^\prime}, \label{tlU}
\end{equation}
where instead of $C_V$ in (\ref{CU}) we should use
\begin{equation}
\tilde C_V = C + \sqrt{\tilde z^\prime} \Sigma \sqrt{\tilde z^\prime}. \label{tcV}
\end{equation}

The self-energy matrix for spin-orbit coupling within the given $l$ subspace is defined as
\begin{equation}
\Sigma_{SO}=\xi_{l\sigma\sigma'}(z)\langle lm\sigma|\mathbf{SL}|lm'\sigma'\rangle
\end{equation}
where the energy dependence of the coupling parameter $\xi_{l\sigma\sigma'}(z)$ comes from the radial wave functions. To second order in $(z-E_\nu)$, the radial function in the LMTO basis is $\phi_{\nu l\sigma}(z)=\phi_{\nu l\sigma}+(z-E_\nu)\phidot_{\nu l\sigma}$. Therefore, we have
\begin{align}
\xi_{l\sigma\sigma'}(z) &= \langle\phi_{\nu l\sigma}(z)|V_{SO}(r) | \phi_{\nu l\sigma'}(z)\rangle\nonumber \\
&= \bigl[\langle \phi_{\nu l \sigma} | V_{SO}(r) | \phi_{\nu l \sigma'} \rangle
+ (z-E_{\nu l\sigma}) \langle \phidot_{\nu l \sigma} | V_{SO}(r) | \phi_{\nu l \sigma'} \rangle
+ (z-E_{\nu l\sigma'}) \langle \phi_{\nu l \sigma} | V_{SO}(r) | \phidot_{\nu l \sigma'} \rangle\nonumber\\
&+ (z-E_{\nu l\sigma})(z-E_{\nu l\sigma'}) \langle \phidot_{\nu l \sigma} | V_{SO}(r) | \phidot_{\nu l \sigma'} \rangle \bigr]/
\sqrt{[1+p_{l\sigma}(z-E_{\nu l\sigma})^2][1+p_{l\sigma'}(z-E_{\nu l\sigma'})^2]} \label{xiz}
\end{align}
where the square roots in the denominator come from the normalization of $\phi_{\nu l\sigma}(z)$.

An approximate form can be obtained by neglecting the terms of order $p(E-E_\nu)^2$ when they appear in the perturbation, i.\ e.\ approximating $\sqrt{\tilde z^\prime}\approx 1$ in (\ref{tcV}) and neglecting the square root in the denominator of (\ref{xiz}). We found that for (Fe$_{1-x}$Co$_x$)$_2$B alloys this approximation captures all the essential physics (see, for example, Fig.\ 1), while the $p(E-E_\nu)^2$ terms only change $K$ by an average of about 10\%.

The above treatment of spin-orbit coupling is extended to CPA with no modifications. Some implementation notes about our CPA code can be found in Ref.\ \onlinecite{S-Fe8N}.

\section{Relation to spin-orbit coupling energy}

When MCA energy $K$ appears in second order in spin-orbit coupling, the anisotropy of the expectation value of the SO operator $\Delta E_{SO}=\langle V_{SO}\rangle_x - \langle V_{SO}\rangle_z$ is approximately equal to $2K$.\cite{AKA14} Here we transcribe this relation in terms of Green's functions.

The single-particle energy is by definition
\begin{equation}
E_{sp} = -\frac{1}{\pi}\im\tr\int z \tilde G_V dz
\end{equation}
The second-order term from perturbation theory is
\begin{equation}
E^{(2)}_{sp} = -\frac{1}{\pi}\im\tr\int z \tilde G_0 \Sigma \tilde G_0 \Sigma \tilde G_0 dz = -\frac{1}{\pi}\im\tr\int z \tilde G^2_0 \Sigma \tilde G_0 \Sigma dz
\end{equation}
The expectation value of $V_{SO}$, with the radial integral included in $\Sigma$, in the same order is
\begin{equation}
\langle V_{SO} \rangle = -\frac{1}{\pi} \im \tr \int \Sigma\tilde G dz \approx -\frac{1}{\pi} \tr \int \Sigma\tilde G_0 \Sigma\tilde G_0 dz .
\end{equation}
An exact quasiparticle Green's function with simple poles of unit residue satisfies $d G/dz = -G^2$. For $\tilde G_0$ this identity is satisfied approximately up to a real term of second order in $(E-E_\nu)$. Thus, setting $d \tilde G_0/dz \approx -\tilde G_0^2$, we find
\begin{equation}
E^{(2)}_{sp} \approx \frac{1}{\pi}\im\tr \int z\frac12\frac{d}{dz} (\tilde G_0\Sigma\tilde G_0\Sigma) dz + ((d\Sigma/dz)) = \frac{\langle V_{SO} \rangle}{2} + ((d\Sigma/dz)) \label{Esp2}
\end{equation}
where the first term was integrated by parts, and $((d\Sigma/dz))$ denotes the terms with the energy derivative of $\Sigma$. For substitutional alloys treated in CPA, the Green's function does not have a purely quasiparticle structure, and in general the relation $d G/dz = -G^2$ is violated due to the energy dependence of the coherent potential. This energy dependence contributes additional terms to the right side of (\ref{Esp2}). However, here we are dealing with relatively weak substitutional Fe/Co disorder, and the electronic structure largely preserves its quasiparticle character. Therefore, we may expect the relation $E^{(2)}_{sp}\approx \langle V_{SO}\rangle/2$ to hold approximately, which is borne out by calculations. Deviations from this relation can also occur near the points of degeneracy close to the Fermi level, where second-order perturbation theory breaks down. As seen in Fig.\ 4, this situation appears near $x=0.8$ where the Fermi level cuts through a pair of flat, nearly degenerate bands. From Fig. 1 we see the largest difference between $K$ and $K_{SO}$ at this point.

\section{Spin-resolved orbital moment anisotropy}

\begin{figure}[htb]
\includegraphics[width=0.6\textwidth]{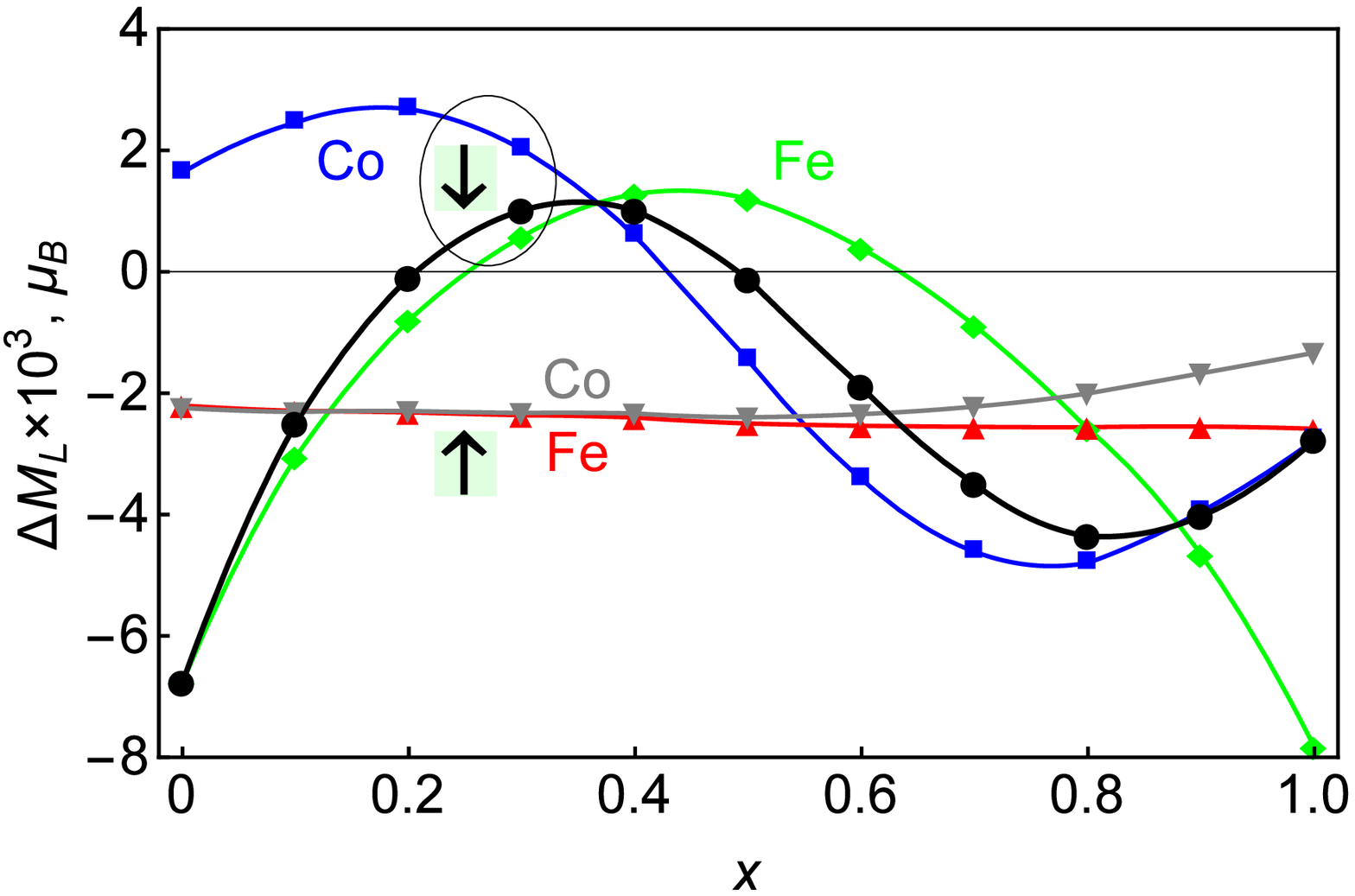}
\caption{Orbital moment anisotropy (OMA) for majority and minority-spin states of Fe and Co. The thick black line shows the concentration-weighted average of Fe and Co contributions from the minority-spin states.}
\label{orbital}
\end{figure}

Orbital moments and their dependence on the direction of magnetization are often discussed in connection with MCA. \cite{Yosida,Streever,Bruno,Laan,S-STOHR,AKA14} In (Fe$_{1-x}$Co$_{x}$)$_{2}$B both spin channels contribute comparably to $K$, and therefore there is no direct relation between $K$ and the anisotropy of the total orbital moment. However, the spin-flip terms are small and change slowly when $x$ is not close to 1 (see Fig.\ 1). Since the energy dependence of the SO parameters is relatively weak, the contributions $K_{\uparrow\uparrow}$ and $K_{\downarrow\downarrow}$ are closely related to the orbital moment anisotropy (OMA) for the states of the corresponding spin. Fig.\ \ref{orbital} shows that the concentration-weighted average of the spin-down OMA behaves similarly to $K_{\downarrow\downarrow}$. The spin-up OMA is almost constant except for the Co-rich end where, as seen in Fig.\ 1, the spin-flip terms also increase substantially. Note that negative OMA for spin-up states corresponds to positive $K_{\uparrow\uparrow}$, but the sign of OMA for spin-down states is the same as that of $K_{\downarrow\downarrow}$. This is because $\langle L_z\rangle_{\downarrow\downarrow}$ appears in $K_{SO}$ with a minus sign, as reflected in Hund's third rule.

The concentration dependence of $K_{\uparrow\uparrow}$ and of the majority-spin OMA is weak because the majority-spin $3d$ bands are fully filled. This feature is typical for Fe, Co, and Ni-based magnetic alloys. Therefore, while $K$ is almost never proportional to the total OMA, their derivatives with respect to the concentration may be strongly correlated in many magnetic $3d$ alloys.

\end{document}